\documentclass[aps,12pt,pra,showpacs,superscriptaddress,preprint]{revtex4}
\usepackage{amsmath}
\usepackage{graphics}
\usepackage{graphicx}
\usepackage{ae,aecompl}
\usepackage{bm}
\usepackage[bookmarks=true,bookmarksopen=true,bookmarksnumbered=true,bookmarksopenlevel=3]{hyperref}

\begin{document}

\newcommand{\bd}{\begin{document}}
\newcommand{\ed}{\end{document}}
\newcommand{\bc}{\begin{center}}
\newcommand{\ec}{\end{center}}
\newcommand{\bfr}{\begin{flushright}}
\newcommand{\efr}{\end{flushright}}
\newcommand{\lt}{\left}
\newcommand{\rt}{\right}
\newcommand{\vs}{\vspace}
\newcommand{\hs}{\hspace}
\newcommand{\beq}{\begin{equation}}
\newcommand{\eeq}{\end{equation}}
\newcommand{\lb}{\linebreak}
\newcommand{\pb}{\pagebreak}
\newcommand{\mb}{\makebox}
\newcommand{\fb}{\framebox}
\newcommand{\mc}{\multicolumn}
\newcommand{\ben}{\begin{enumerate}}
\newcommand{\een}{\end{enumerate}}
\newcommand{\bit}{\begin{itemize}}
\newcommand{\eit}{\end{itemize}}
\newcommand{\un}{\underline}
\newcommand{\lefq}{\lefteqn}
\newcommand{\ba}{\begin{array}}
\newcommand{\ea}{\end{array}}
\newcommand{\beqa}{\begin{eqnarray}}
\newcommand{\eeqa}{\end{eqnarray}}
\newcommand{\beqas}{\begin{eqnarray*}}
\newcommand{\eeqas}{\end{eqnarray*}}
\newcommand{\bfg}{\begin{figure}}
\newcommand{\efg}{\end{figure}}
\newcommand{\bds}{\begin{displaymath}}
\newcommand{\eds}{\end{displaymath}}
\newcommand{\btb}{\begin{tabbing}}
\newcommand{\etb}{\end{tabbing}}
\newcommand{\para}{\parallel}
\newcommand{\pad}{\partial}
\newcommand{\nn}{\nonumber}
\newcommand{\la}{\leftarrow}
\newcommand{\ra}{\rightarrow}
\newcommand{\lgla}{\longleftarrow}
\newcommand{\lgra}{\longrightarrow}
\newcommand{\La}{\Leftarrow}\newcommand{\Ra}{\Rightarrow}
\newcommand{\Lra}{\Leftrightarrow}
\newcommand{\Lgla}{\Longleftarrow}
\newcommand{\Lgra}{\Longrightarrow}
\newcommand{\lan}{\langle}
\newcommand{\ran}{\rangle}
\renewcommand{\a}{\alpha}
\renewcommand{\b}{\beta}
\newcommand{\g}{\gamma}
\newcommand{\G}{\Gamma}
\renewcommand{\d}{\delta}
\newcommand{\eps}{\epsilon}
\newcommand{\Th}{\Theta}
\newcommand{\s}{\sigma}
\newcommand{\lam}{\lambda}
\newcommand{\D}{\Delta}
\newcommand{\vare}{\varepsilon}
\newcommand{\pr}{\prime}
\newcommand{\ro}{\rho}
\newcommand{\nab}{\nabla}
\newcommand{\m}{\mu}
\newcommand{\n}{\nu}
\newcommand{\Sg}{\Sigma}
\newcommand{\p}{\pi}
\newcommand{\R}{I\!\!R}
\newcommand{\om}{\omega}
\newcommand{\Om}{\Omega}
\newcommand{\ze}{\zeta}
\newcommand{\vart}{\vartheta}
\newcommand{\tri}{\triangle}
\newcommand{\f}{\frac}
\newcommand{\iny}{\infty}
\newcommand{\pro}{\propto}

\title{
Bound state in continuum like solutions in one dimensional hetero-structures}

\author{\textsc{O.~Panella}}
\affiliation{Istituto Nazionale di Fisica Nucleare, Sezione di Perugia, Via A.~Pascoli, I-06123 Perugia, Italy}
\email{orlando.panella@pg.infn.it ({\bf Corresponding Author}). }
\altaffiliation[Permanent Address: ]{INFN Sezione di Perugia, Via A Pascoli, I-06123, Perugia, Italy. Phone: +39 075 5852762; Fax: +39 075 44666.}

\author{\textsc{P.~Roy}}
\affiliation{Physics and Applied Mathematics Unit\\
Indian Statistical Institute  \\ Kolkata,
India}

%\date{December 1, 2008}
\date{\today}

\begin{abstract}
We examine the one dimensional Dirac equation with modulated or position dependent velocity. In particular, it is shown that using suitable velocity profiles it is possible to create bound state in continuum (BIC) like, as well as, discrete energy bound state solutions.
\end{abstract}


\maketitle

\section{Introduction} 
Since the rise of graphene \cite{geim,geim1} there have been a lot of studies on the application of Dirac equation in condensed matter systems.
In such systems the effective low energy model of the quasi particles is described by the Dirac equation with a Fermi velocity $v_F$ which is $1/300$ times the velocity of light. The peculiar property of charge carriers in graphene is that they behave as relativistic massless fermions. Subsequently, there have been numerous studies on various theoretical as well as experimental aspects of graphene. In particular, confinement of the quasi particles using various types of magnetic and electrostatic fields have been studied \cite{confine}. Other types of physically interesting states like the quasi bound states \cite{matulis} have also been investigated. Here our interest lies in finding \emph{bound states in the continuum} (BIC states) as well as discrete energy states in one dimensional heterostructures governed by the Dirac equation. The BIC are bound states embedded in the continuum \cite{von}. Such states have a long history \cite{sukhatme} and some years back the possibility of detecting BIC in semiconductor heterostructures has also been pointed out \cite{capasso}. Later, BIC has been detected in photonic systems \cite{marina,bulga}. The possibility of creating BIC in graphene quantum dot structures has also been studied \cite{bic}.

On the other hand, there have been a number of studies on observable effects of modulated velocity or position dependent velocity in one dimensional heterostructures as well as graphene \cite{vel,vel1,raoux,concha}. In such a scenario, the Fermi velocity is different in parts of the material. Backscattering in such systems described by one dimensional Dirac equation has recently been investigated thoroughly in ref \cite{vel}. Here our objective is to examine the$(1+1)$ dimensional Dirac equation with position dependent velocity. To be more specific, we shall investigate the existence of BIC like states as well as discrete energy bound states in the presence of velocity profiles which continuously depends on the position.

\section{The Model} 
The Hamiltonian of a quasi-particle is of the form 
\beq
H=v_F\s_xp_x\label{h1}
\eeq
where $\s_x$ denotes the Pauli matrix. Now in the case of a heterostructure, it is quite possible that velocity may be different in two regions of the material. In other words, the velocity becomes space dependent. However, this can not be achieved by simply replacing $v_F$ by $v_F(x)$ in Eq.(\ref{h1}) since it would render the Hamiltonian non Hermitian. To keep the Hamiltonian Hermitian if the velocity $v_F$ becomes position dependent it is necessary to write it in the following way \cite{vel}
\beq
H=\sqrt{v_F(x)}\s_xp_x\sqrt{v_F(x)}\label{h2}
\eeq
It is easy to see that in case the velocity is a constant, the Hamiltonian (\ref{h2}) reduces to the form (\ref{h1}). The Hamiltonian in Eq.~(\ref{h2}) being hermitian we can also easily derive the associated continuity equation which expresses the conservation of probability:
\begin{equation}
\frac{\partial \rho}{\partial t} + \frac{\partial j_x}{\partial x} =0
\end{equation}
where the current associated to the probability density $\rho= \psi^{\dagger}\psi=
\psi_1^*\psi_1+\psi_2^*\psi_2 $ is given by:
\begin{equation}
j_x = \sqrt{v_F} \psi_1^* \, \sqrt{v_F} \psi_2 + \text{c.c.}
\end{equation}
The Hamiltonian in Eq.~(\ref{h2}) operates on two component spinors $\psi(x)=(\psi_1(x)~,~\psi_2(x))^T$ and the coupled equations for the components can now be written as
\beq
\sqrt{v_F(x)}\, p_x\left[\sqrt{v_F(x)}\,\psi_2(x)\right]=E\,\psi_1(x)\label{eigen1}
\eeq
\beq
\sqrt{v_F(x)}\, p_x\left[\sqrt{v_F(x)}\,\psi_1(x) \right]=E\, \psi_2(x)\label{eigen2}
\eeq
Now introducing an auxiliary spinor $u_{1,2}(x)=\sqrt{v_F(x)}\, \psi_{1,2} (x)$, the above pair of equations may be written as
\beq
v_Fp_xu_1=Eu_2\label{intert1}
\eeq
\beq
v_Fp_xu_2=Eu_1\label{intert2}
\eeq
Let us now, for example, obtain the equation for the component $u_1$ by eliminating $u_2$ from Eqs.(\ref{intert1}) and (\ref{intert2}) :
\beq
v_F^2\f{d^2u_1}{dx^2}+v_Fv_F^\prime \f{du_1}{dx}=\eps^2 u_1,~~~~\eps^2=E^2/\hbar^2\label{u1}
\eeq
where the prime indicates differentiation w.r.t $x$. Let us now apply a further transformation involving the independent variable as
\beq
z=\int^x \f{1}{v_F(t)}dt\label{ztr}
\eeq
and obtain from Eq.(\ref{u1})
\beq
\f{d^2u_1}{dz^2}+\eps^2u_1=0\label{u12}
\eeq
Eq.(\ref{u12}) looks like a free particle equation. However, whether or not it is a free particle equation depends on the range of $z$ which in turn depends on $v_F(x)$. We shall now examine the scenario with a couple of different choices of $v_F(x)$.

\section{BIC like solutions}
Here we shall examine the existence of BIC like solutions of Eq.(\ref{u12}) using different choices of the velocity $v_F(x)$.

\noindent{\bf Example 1.}

Let us first consider 
\beq
v_F(x)=v_0\cosh^2(\a x)\label{vf1},~~~~v_0~\text{is~a~constant} >0
\eeq
It follows that for $\a\ra 0$ we recover the constant velocity scenario. Then from (\ref{ztr}) we find $\displaystyle -\f{1}{\a v_0}\leq z=\f{1}{\a v_0}\tanh~\a x\leq \f{1}{\a v_0}$ and consequently Eq.({\ref{u12}) does not represent a free particle. However, we choose a formal or unphysical solution \footnote{This solution does not satisfy the boundary condition at $\pm \f{1}{\a v_0}$} of Eq.(\ref{u12}), namely,
\beq
u_1=\sin\,(\eps z)
\eeq
Then the component $\psi_1(x)$ of the original problem is given by
\beq
\psi_1(x)=\text{sech}(\a x)~\sin\left(\f{\eps}{\a v_0}~ \tanh\a x\right)\label{psi1}
\eeq
%where 
%\beq
%N_1= \left[\f{1}{\a}-\f{v_0\sin(2\eps/\a v_0)}{2\eps}\right]^{-1/2}
%\eeq
Clearly, the second part of $\psi_1(x)$ is an oscillatory function bounded between $\pm 1$ while the first part is a function decreasing at both ends. Consequently, $\psi_1(x)$ as given by Eq.(\ref{psi1}) is a normalizable solution \emph{for any value of the energy} $E$ and we conclude that Eq.(\ref{psi1}) represents a BIC like solution. 
Now using the relation (\ref{intert1}) the second component of the spinor can be found to be
\beq
\psi_2(x)=-i~\text{sech}(\a x)~\cos\left[\f{\eps}{\a v_0}\,\tanh(\a x)\right]\label{psi2}
\eeq
%where
%\beq
%N_2=\left[\f{1}{\a}+\f{v_0\sin(2\eps/\a v_0)}{2\eps}\right]^{-1/2}
%\eeq 
Clearly both components $\psi_{1,2}$ are square integrable functions. One can indeed exactly evaluate the normalization constant $N$ for the complete Dirac spinor. Choosing to normalize to 1 the total probability density $\rho=\psi^{\dagger} \psi$  we have:
\begin{eqnarray}
\int_{-\infty}^{+\infty} dx \rho(x)& =&  \int_{-\infty}^{+\infty} \, dx\,\left[ |\psi_1(x)|^2 +|\psi_2(x)|^2\right] \nonumber \\&=& \frac{2}{\alpha}N^2\phantom{xxxxx}
\end{eqnarray}
and the normalization constant is given by $N=\sqrt{\alpha/2}$.
%\beq
%N=\left[\f{2}{\a}
%-\f{v_0\sin(2\eps/\a v_0)}{2\eps}
%\right]^{-1/2}\label{b}
%\eeq
We note that the two components of  our spinor solution contribute differently,  in general, to the total probability density.  In any case, both components $\psi_{1,2}(x)$  and consequently the Dirac spinor 
$\psi=(\psi_1, \psi_2)^T$
%$\left(\ba {cc} \psi_1\\ \psi_2\ea\right)$ 
represent a BIC like solution.
In Fig.~\ref{figex1} we have drawn a plot of $\psi_1(x)$ (left panel) and  the normalized probability density $\rho(x) = \psi^{\dagger}\psi=\psi_1^*\psi_1+\psi_2^*\psi_2$ (right panel) for a particular set of parameters in arbitrary units (see figure caption). It can be seen that the wavefunction oscillates with decreasing amplitude and eventually goes to zero.  Also, the plot of the probability density (Fig.~\ref{figex1} right panel) shows that increasing the parameter $\alpha$ increases the localization of the BIC state. 

We can also easily check that the states described by Eqs.~(\ref{psi1},\ref{psi2}) have a vanishing probability current density as it is expected for a bound state:
\begin{equation}
j_x= N \sqrt{v_0} \left[\,(i-i )\, \sin\left(\f{\eps}{\a v_0}~ \tanh\a x \right) \,\cos\left(\f{\eps}{\a v_0}~ \tanh\a x \right)\,\right] = 0\,. 
\end{equation}
\begin{figure}
\scalebox{1.1}{\includegraphics{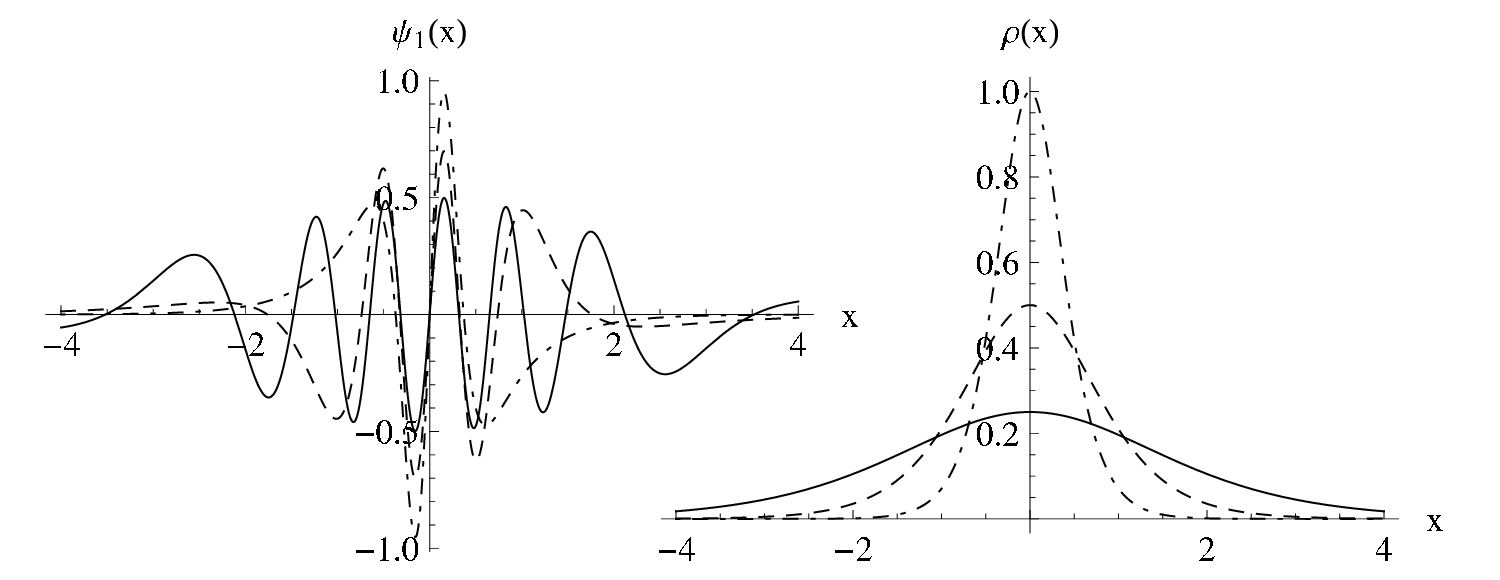}}
  \caption{(Left panel) Plot of the component $\psi_1(x)$, see Eq.~\ref{psi1} for the following set of values  for the parameters: $v_0=1, \eps=10, \a=0.5\, (\text{solid curve}), \a=1\, (\text{dashed curve)}, \a=2\, (\text{dot-dashed curve}) $ (arbitrary units).
  (Right panel)  Plot of the (normalized) probability density $\rho(x)$ for the same values of the parameters as in the left panel. We clearly see that increasing the parameter $\a$ increases the localization of the BIC state. }
  \label{figex1}
\end{figure}

\pb
{\bf Example 2.}

Here we choose
\beq
v_F(x)=v_0(1+\a x^2)^2,~~~~\a>0
\eeq
Then it follows that 
\beq
-\f{\pi}{2\sqrt{\a}v_0}\leq z = \f{1}{2v_0}\left[\f{x}{(1+\a x^2)}+\f{1}{\sqrt{\a}}\arctan\sqrt{\a}x\right]\leq \f{\pi}{2\sqrt{\a}v_0}
\eeq
Now proceeding as before the solutions $\psi_{1,2}(x)$ are found to be
\beq
\psi_1(x)= \, \f{1}{1+\a x^2}\,\sin\left[\f{\eps}{2v_0}\left(\f{x}{1+\a x^2}+\f{1}{\sqrt{\a}}\arctan\sqrt{\a}x\right)\right]\label{wf2}
\eeq
\beq
\psi_2(x)=-i\, \f{1}{1+\a x^2}\,\cos\left[\f{\eps}{2v_0}\left(\f{x}{1+\a x^2}+\f{1}{\sqrt{\a}}\arctan\sqrt{\a}x\right)\right]\label{wf3}
\eeq
As in the previous example both the components are square integrable functions and they contribute differently to the total probability density $\rho=\psi^{\dagger}\psi$.  Normalizing to 1 the total probability density we find:
\begin{eqnarray}
\int_{-\infty}^{+\infty} dx \rho(x)& =&  \int_{-\infty}^{+\infty} \, dx\,\left[ |\psi_1(x)|^2 +|\psi_2(x)|^2\right] \nonumber \\
&=&\frac{\pi}{\sqrt{\alpha}}N^2
\end{eqnarray}
where $N=\sqrt{\sqrt{\alpha}/\pi}$.  

%\beq
%N=\left[\f{\pi}{2\sqrt{\a}}-\f{v_0^2\, \sin\left(\f{\pi\eps}{2v_0^2\sqrt{\a}}\right)}{\eps}\right]^{-1/2}
%\eeq
It is readily seen that the solutions (\ref{wf2}) and (\ref{wf3}) are valid for any value of $\eps$, and hence \emph{for any value of  the energy eigenvalue} $E$. 
In Fig.~\ref{figex2} we have plotted the component wave function $\psi_1$ of Eq.~(\ref{wf2}) (left panel) and the normalized probability density (right panel) for the same set of parameters (in arbitrary units) as in Example 1. As in the first example, it is also seen here that the wave function oscillates with non constant amplitude before dying out. The probability density shown in the right panel again shows that  with larger values of the parameter $\alpha$ there is stronger localization. 

From these two examples it is clear that one may choose many different profiles for $v_F(x)$ which would produce a BIC like solution.
\begin{figure}
\scalebox{1.1}{\includegraphics{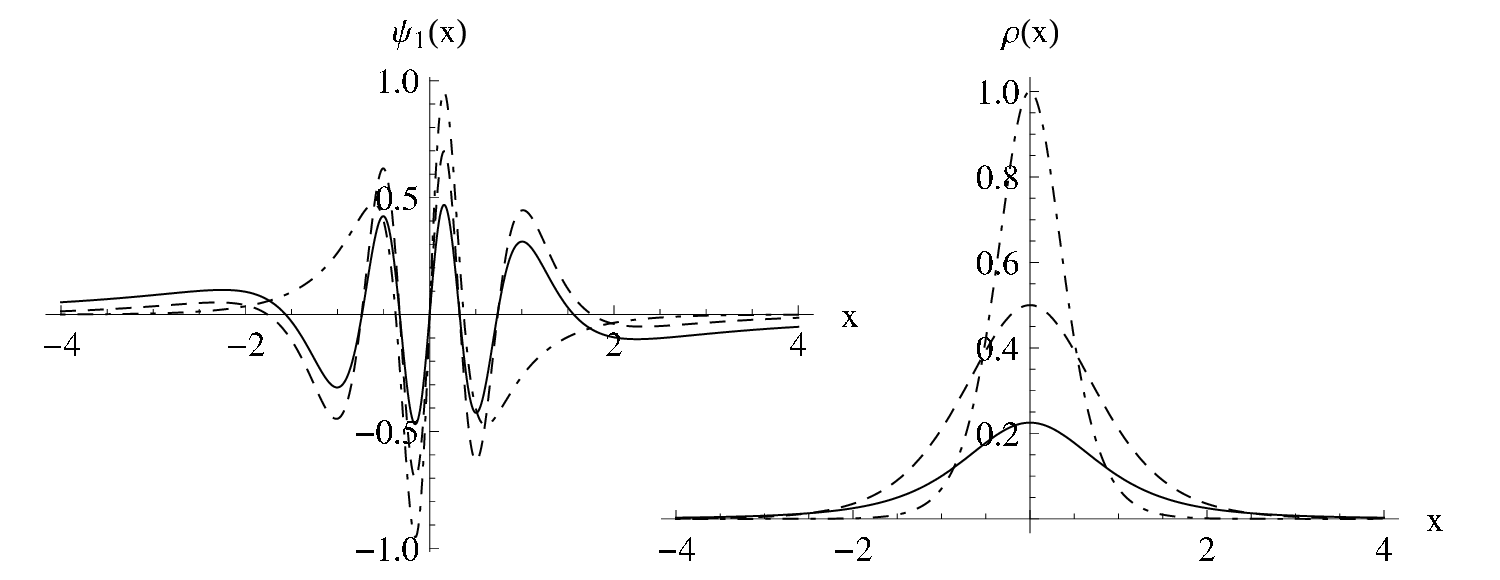}}
  \caption{(Left panel) Plot of the component $\psi_1(x)$, see Eq.~\ref{wf2} relative to Example 2, for the following set of values  for the parameters: $v_0=1, \eps=10, \a=0.5\, (\text{solid curve}), \a=1\, (\text{dashed curve)}, \a=2\, (\text{dot-dashed curve}) $ (arbitrary units).
(Right panel)  Plot of the (normalized) probability density $\rho(x)$ for the same values of the parameters as in the left panel. We clearly see also in this second example that increasing the parameter $\a$ increases the localization of the BIC state. }
  \label{figex2}
\end{figure}

\subsection{Bound state solutions}
In both the examples above we have considered unphysical solutions of Eq.(\ref{u12}). It now remains to be seen what happens when a physical solution i.e., a solution which vanishes at the boundaries is chosen. Here we shall only examine Example 1 in detail. In this case the problem is clearly that of a particle trapped in an infinite square well potential. The solutions of this problem are well known and are given by
%\beq
%u_1(z)=\sqrt{\a v_0}\, \sin\left[\f{\pi\a v_0n}{2}(z+\f{1}{\a v_0})\right],~~~~\eps_n^2=\f{\a^2v_0^2\pi^2n^2}{4},~~~~n=1,2,3,......
%\eeq
%Hence the solutions of the original problem are given by
\beq
u_1(z)= \sin\left[\f{\pi\a v_0n}{2}(z+\f{1}{\a v_0})\right],~~~~\eps_n^2=\f{\a^2v_0^2\pi^2n^2}{4},~~~~n=1,2,3,......
\eeq
Hence the solutions of the original problem are given by
\beq
\psi_{1,n}(x)= \text{sech}(\a x)~\sin\left[\f{\pi n}{2}(\tanh\a x+1)\right],~~~~E_{1,n}=\f{\hbar\a v_0\pi n}{2}\label{f}
\eeq
One may easily check that the solutions (\ref{f}) are normalizable. Now again using relation (\ref{intert1}) we obtain
\beq
\psi_{2,n}(x)=-i~\text{sech}(\a x)~\cos\left[\f{\pi n}{2}(\tanh\a x+1)\right],~~~~E_{2,n}=\f{\hbar\a v_0\pi n}{2}
\eeq
It may be observed that the solution $\psi_{2,n}(x)$ represents the {\it same energy level} as $\psi_{1,n}(x)$. Therefore the complete solution is given by
\beq
\psi_n(x) = \sqrt{\f{\a}{2}}~\text{sech}{\a x}\left(\ba{cc}\phantom{-i}~\sin\left[\displaystyle\f{\pi n}{2}(\tanh\a x+1)\right]\\-i~\cos\left[\displaystyle\f{\pi n}{2}(\tanh\a x+1)\right]\ea\right),~~~~E_n=\f{\hbar\a v_0\pi n}{2}
\eeq
So, in conclusion, whether or not the solutions of the original problem would represent a BIC or bound state depends on whether we choose an unphysical or a physical solution of the transformed equation given by  Eq.(\ref{u12}).

%\psi_{1,n}(x)=\sqrt{\f{\a}{v_0}}\, \text{sech}(\a x)~\sin\left[\f{\pi n}{2}(\tanh\a x+1)\right],~~~~E_{1,n}=\f{\hbar\a v_0\pi n}{2}\label{f}
%\eeq
%One may easily check that the solutions (\ref{f}) are normalizable. Now again using relation (\ref{intert1}) we obtain
%\beq
%\psi_{2,n}(x)=-i\sqrt{\f{\a}{v_0}}~\text{sech}(\a x)~\cos\left[\f{\pi n}{2}(\tanh\a x+1)\right],~~~~E_{2,n}=\f{\hbar\a v_0\pi n}{2}
%\eeq
%It is interesting to observe that the solution $\psi_2(x)$ represents the same energy level as $\psi_1(x)$. So, in conclusion, whether or not the solutions of the original problem would represent a BIC or bound state depend on whether we choose an unphysical or a physical solution of the transformed equation given by  Eq.(\ref{u12}).

\section{Conclusion}
Here we have examined the possibility of obtaining BIC as well as bound state solutions within the framework of one dimensional Dirac equation with a velocity which depends continuously on position. In particular, it has been shown that such solutions do exist for suitable velocity profiles. We feel it would be of interest to investigate BIC in the presence of a position dependent mass or a scalar potential as well looking for similar solutions in two dimensions. 

\begin{acknowledgments}
%{\bf Acknowledgement}
One of us (P.~R.) wishes to thank INFN Sezione di Perugia for supporting a visit during which this work was carried out.  The Physics Department of the University of Perugia is also acknowledged
for the kind hospitality.
\end{acknowledgments}

\ed